`

# Phase Transitions Driven by Quasi-Particle Interactions – II

Norman H. March*, Richard H. Squire$^\xi$,

\* Department of Physics, University of Antwerp (RUCA), Groenborgerlaan 171,
B-2020 Antwerp, Belgium
and
Oxford University, Oxford, England

$^\xi$ Department of Chemistry, West Virginia University
Institute of Technology
Montgomery, WV 25303, USA

## ABSTRACT

Quasi-particles and collective effects may have seemed exotic when first proposed in the 1930's, but their status has blossomed with their confirmation by today's sophisticated experiment techniques. Evidence has accumulated about the interactions of, say, magnons and rotons and with each other and also other quasi-particles. We briefly review the conjectures of their existence necessary to provide quantitative agreement with experiment which in the early period was their only raison d'être. Phase transitions in the Anderson model, the Kondo effect, roton-roton interactions, and highly correlated systems such as helium-4, the Quantum Hall Effect, and BEC condensates are discussed. Some insulator and superconductor theories seem to suggest collective interactions of several quasi-particles may be necessary to explain the behavior. We conclude with brief discussions of the possibility of using the Grüneisen parameter to detect quantum critical points and some background on bound states emerging from the continuum. Finally, we present a summary and conclusions and also discuss possible future directions.

**1. Introduction and Context.** The literature on collective effects and quasi-particles can be quite difficult to explore as some early studies in condensed matter may not have explicitly use the term(s) and/or they may be in a different field. On the other hand one only needs to look at the "multi-discipline" impact of BCS theory on both condensed matter and nuclear theory [1]. Introduction to these concepts begins with Bloch, Peierls and Landau, three of the earliest: Bloch introduced low energy excitations (spin waves) in ferromagnets [2] and Peierls suggest a possible form for superconductivity being discussed even today [3]. Landau followed with a low energy collective excitation in liquid helium-4, the roton, which he proposed had an influence on its viscosity [4]. In fact he demonstrated that roton-roton scattering needed to be taken into account [5] and it could be done in a quantitative manner with a phenomenological "quantum hydrodynamic" approach. This might be one of the first detailed discussions of quasi-particle (QP) interactions; others followed using "first principles" (see Parry et al for an early review [6], refs 7 -25 to put the early work more fully into perspective). Dyson discussed magnon-magnon interactions as a result of two spin waves binding together and traveling through the lattice [7]   Later, Landau developed one of the standard models, Fermi liquid theory (FLT) [8] where the elementary excitations are called quasi-particles which are composite particles with a lifetime which approaches infinity as its energy approaches the Fermi level. In addition the scattering rates scale as $(\varepsilon_1 - \varepsilon_F)^2$. Indeed, the BCS theory assumes that the properties of the normal state are non-interacting quasi-particles with infinite lifetimes [1].

Our discussion may mix fermion and boson quasi-particles, but this can occur naturally. For example in BCS theory one can derive a Cooper pair from a two body interaction which appears to be an entirely fermionic process in first quantization. Alternatively, if we use second quantization (field theory), the process appears differently (see [9, 10] for discussions). In addition dimensionality, magnetic field, pressure, doping, and other considerations may affect the nature of quasi-particles, their interactions and/or collective effects.

It was pointed out by one of the authors and coworkers [11] that the transition temperatures $T_t$ for different phase transitions can have a "shape"

$$k_B T_t = E_{char} \exp\left(-const/\lambda\right) \qquad (1.1)$$

where $\lambda$ is a measure of a quasi-particle interaction based on experimental and phenomenological arguments. The quasi-particles considered by the previous authors were phonon-phonon, electron-phonon, and magnon-magnon interactions pertinent to solid-liquid, superconducting and Peierls, and ferromagnetic-paramagnetic transitions, respectively. For example, a relationship was proposed for the melting of crystalline metals at temperature $k_B T_m$ where

$$E_{char} = const \cdot B \cdot \Omega$$

with B, the bulk modulus and $\Omega$ the atomic volume,

$$k_B T_m (overall) = (\sim 0.3) \exp\left(-const/\lambda\right) = (\sim 0.3) \exp(-0.5\gamma)$$

Here, the pre-exponential 0.3 is semi-empirical, and $\gamma$ is the Grüneisen parameter, $\gamma = \alpha/C_P$, the ratio of the thermal expansivity $\left(\alpha = \frac{1}{V}\frac{\partial V}{\partial T}\right)$ to the molar specific heat, $C_P$, experimentally measuring anharmonicity. In Section 6 we discuss an interesting expansion of the Grüneisen parameter as applies to possibly detecting quantum critical points.

First we discuss additional models (relative to [11], with some comparison of some seemingly similar quasi-particles in different contexts. This is followed by a discussion of the Grüneisen parameter as might be applied to detect quantum critical points and a discussion of the bound states that may result from quasi-particle interactions. The various examples are summarized in a Table. A summary is given, together with conclusions and proposals for future work.

## 2. Magnetic Examples

**A. The Anderson Model [12].** The basic Anderson model is composed of two terms, an "atomic limit" $H_{atomic}$ which describes an isolated magnetic ion such as an f state (Kramer's doublet) and $H_{resonance}$ describing the hybridization of the Bloch conduction waves with the localized f electrons,

$$H = H_{atomic} + H_{resonance}$$

where $H_{atomic} = E_f n_f + U n_{f\uparrow} n_{f\downarrow}$ and $H_{resonance} = \sum_{k,\sigma} \varepsilon_k n_{k\sigma} + \sum_{k,\sigma} V(k)\left[c_{k\sigma}^\dagger f_\sigma + f_\sigma^\dagger c_{k\sigma}\right]$ (2.1)

One can approach the model from the atomic limit where $V(k) = 0$ and the hybridization is increased or the adiabatic view where $U$, the cost of double occupancy of an orbital, goes from $0 \to U$. The merger of the two approaches is the recognition that quantum spin fluctuations cause the local moment to tunnel on a timescale $\tau_{sf}$ between two degenerate configurations,

$$e_\uparrow^- + f_\downarrow^1 \rightleftharpoons e_\downarrow^- + f_\uparrow^1$$

We discuss the importance of timescale in the Kondo model below, so we finish with the temperature $T_{SA}$ (symmetric Anderson) relationship

$$T_{SA} = \sqrt{\frac{2U\Delta}{\pi^2}} \exp\left(-\frac{\pi U}{8\Delta}\right) \qquad (2.2)$$

This temperature is a crossover from a low temperature paramagnetic susceptibility $\chi \sim 1/T_{SA}$ to Curie law susceptibility $\chi \sim 1/T$ at higher temperature.

**B. The Kondo Model [13].** Using the timescale $\tau_{sf}$ from section A above, the fluctuations between the degenerate configurations below a temperature corresponding to thermal energy $k_B T$ are of the order of the characteristic tunneling rate $\hbar/\tau_{sf}$ which allows formation of a paramagnetic state with a Fermi-liquid resonance, the width of the resonance being determined by the Kondo energy $k_B T_K \sim \hbar/\tau_{sf}$.

$$T_k = (\Gamma\mu)^{1/2} \exp\left(\frac{-2\Gamma}{\pi|\varepsilon_d|}\right) \qquad (2.3)$$

Our discussion does not due justice to the intriguing connections between "heavy electron" theory, superconductivity and the Kondo effect [14]. It should be noted that a key connection between the Anderson and Kondo models was established by Schrieffer and Wolff [15] where $\Gamma$ is the scattering rate between conduction and localized orbitals and U is is is the energy cost for a doubly occupied orbital.

$$J = -\frac{8\Gamma}{\pi U}$$

The Kondo effect has been used to create a single-electron transistor [16, 17].

**C. Curie Temperature**

Villain and Bak have studied an anisotropic Ising model (the so called "anisotropic next nearest neighbor Ising" – ANNNI) containing interactions of stripes of positive and negative spins [18]. This problem is rich in structure as exemplified by the concept of a "floating phase" with a continuously varying wave vector [19, 20, 21, 22]. Using reasonable approximations result in a Gaussian/harmonic approximation providing a connection to both a magnon model in earlier work [7] and later to Kosterlitz-Thouless (KT) vortices [23]. The harmonic approach as discussed by deGennes [24] can be related to the Kohn anomaly [25] and thus through the analogous structure of magnons and phonons to an implicit formula for the Curie temperature,

$$\frac{k_B T_c}{J} e^{-2J/k_B T_c} = Const(J) \qquad (2.4)$$

where J is the strength of the exchange coupling. In other situations the magnon may simply provide a background for the topological vortex of the KT theory,

$$k_B T_c \approx \pi J \qquad (2.5)$$

Villain's attempts to simplify the resulting partition function resulted in the "Villain model" [26], while Jose′ and others [27] managed to sum the partition functions so they can be expressed in terms of quantum numbers identifying the vortices. Mattis has written a nice summary as well as a discussion of magnons [28].

**3. Roton-roton interaction.** Considerable work has been completed since the review of Parry and ter Haar [6]. This discussion will contain four subsections: A) helium, B) the QHE, C) composite fermions, and D) Bose-Einstein Condensates (BEC).

**A. Helium**. Landau's roton proposal some seventy years ago that the low energy excitation spectrum consisted of low energy phonon excitations and rotons at higher energies offered an explanation of some of the peculiar thermodynamic properties such as viscosity of liquid helium. Later, Landau and Khlatnikov studied the possibility that roton-roton interactions contributed to the liquid helium viscosity; their work used a delta function interaction in their "hydrodynamic formulation" and they were surprisingly accurate. As indicated in Parry's review, Feynman and Cohen performed several first principles calculations which suggested that the roton-roton interaction was likely to be complicated and depend on the orientation of the rotons. While their spectrum qualitatively agreed with Landau's, it was quantitatively high in energy. Greytak et al [29] used Raman scattering to study helium rotons pairs. Their experimental results revealed only a single peak near twice the single roton energy. Zawadowski, Ruvalds and Solano (ZRS) [30] assumed that the roton-roton interaction is attractive and suggested a bound state below the two roton continuum at energy $\varepsilon = 2\Delta_0 - E_b$, $E_b$ being the binding energy. Graytak then confirmed the binding energy and the D-type angular momentum of the bound state using extremely precise measurements.

ZRS then hybridized the single particle and two-roton spectrum with roton-roton interactions which resulted in a very strong distortion of the single particle spectrum; the single particle state splitting into two branches, consistent with neutron data after the inclusion of a finite roton lifetime. It is only with the strong hybridization that their calculated spectrum shows very good agreement with the experiment.

ZRS recognize that a bound roton pair can be formed below the continuum in a manner analogous to the BCS theory, i.e. a bound state will appear from the continuum no matter how small the attractive coupling with a binding energy, $E_b$,

$$E_b = 2D \exp\left(-\frac{1}{2|\rho_0(K)g_4|}\right) \qquad (3.1)$$

and D is the cutoff energy, and $g_3(T) = g_4 \left[N_0(T)\right]^{1/2}$ ($g_3$ is the interaction between excitations one- and two-particle states, and $N_0$ is the temperature dependent number density) and $\rho_0$ is the joint density of states. Later Pines et al [31] used a pseudo potential method based on the experimental helium potential curve and calculated two-roton bound states, roton-liquid parameters, and roton lifetimes, as well as information about the hybridization of the two-roton bound state. In their work there is excellent agreement between theory and experiment for both excitations of higher and lower energy.

**B. Quantum Hall Rotons.** Thirty years ago von Klitzing [32] subjected a two-dimensional (2D) electron system to very low temperatures and high magnetic fields and observed that the conductivity $\sigma$ was precisely quantized as $\sigma = \upsilon \frac{e^2}{h}$ where e is the elementary charge on an electron, h is Planck's constant, and $\upsilon$ is called a "filling factor". The ratio of the transverse potential difference to the total current plotted against the magnetic field showed flat regions where $\upsilon$ was an integer to high precision; the results with did not vary from sample to sample. Landau's 2D model of an electron gas and the plateaus mentioned above are called Landau levels. The Hall conductance was identified as a topological invariant [33].

Two years later Tsui et al [34] was discovered that $\upsilon$ did not need be an integer, but could assume fractional values (FQHE) $\upsilon = \dfrac{1}{2n+1}$. Laughlin then created a wave function for the lowest Landau level utilizing an angular momentum barrier to keep electrons separated [35].

In the development of effective theories to describe the FQHE Girvin, MacDonald, and Platzman (GMP) [36] modified Feynman's excitation theory for helium superfluid to describe the collective excitation gap with the analogy that the Hall resistance is almost dissipation-less much like a superfluid. A brief recounting is given here so the comparison with rotons is clearer. According to Feynman, if the exact ground state wave function is known, the density wave excited state variational wave function can be written as (at wave vector $\vec{k}$ ),

$$\Phi_{\vec{k}} = \frac{1}{\sqrt{N}} \rho_{\vec{k}} \psi \tag{3.2}$$

N being the number of particles and $\rho_{\vec{k}}$, the Fourier transform of the density operator

$$\rho_{\vec{k}} = \sum_{j=1}^{N} e^{i\vec{k}\cdot\vec{r}_j} \tag{3.3}$$

Note that $\Phi_{\vec{k}}$ is orthogonal to $\psi$ which contains considerable correlations by its very nature, i.e. the ground state wave function. Feynman states that an excited state configuration has three criteria to be met: 1) it must be similar to the ground state; 2) it must have a large density $\rho_{\vec{k}}$ ; and 3) $\rho_{\vec{k}}$ must have a significant density modulation at $\vec{k}$. The result is the Bijl-Feynman equation:

$$\Delta(k) = \frac{f(k)}{s(k)} \tag{3.4}$$

with the norm of the excited state being

$$f(k) = \frac{\langle \psi | \rho_k^\dagger (H - E_0) \rho_k | \psi \rangle}{N} \tag{3.5}$$

($H$ is the Hamiltonian and $E_0$, the ground state). The utility of eq (3.4) is that the collective mode, a dynamical quantity, is expressed in terms of static properties of the ground state since $f(k)$ is the oscillator strength expressed by the sum rule

$$f(k) = \frac{\hbar^2 k^2}{2m} \tag{3.6}$$

and $s(k)$ is the static structure factor determined by neutron scattering experiments in helium. Essentially, eq. (3.6) is the collective mode energy $\Delta(k)$ is the single particle energy at a momentum divided by a particle correlation renormalization. GMP's theory gives an excellent quantitative fit to experiment, and proved especially interesting when quantitative evidence for off-diagonal long range order (ODLRO) in Quantum Hall States was found [37]. ODLRO is a characteristic of BEC, and superconductivity where the condensate density $n_0$ in an interacting Bose gas has a macroscopic eigenvalue of the one particle density matrix [38, 39],

$$n_0 = \lim_{|r-r'|\to\infty} \rho_1(\vec{r}_1 - \vec{r}_1')$$

The theory has evolved to a point where inter-Landau level modes have been named magneto-plasmons and intra-Landau level modes are called the magneto-phonons and the minimum away from k = 0 is a magneto-roton. Recently studies in FQHE have led to the first experimental evidence for rotons [40, 41, 42]. It seems that the Quantum Hall Effect could be thought of as a result of a transition of an electron liquid to an electron superfluid by creating a two dimensional system in high magnetic fields at low temperatures with quasi-particle participation. General discussions are available [43].

**C. Composite Fermions.** Another approach to the IQHE is through composite Fermion (CF) analysis pioneered by Jain [44] which is amenable to numerical studies. Recent by analytical work by Ghosh and Baskaran [45] and numerical work by Park and Jain [46] has illustrated that the magneto roton zero momentum energy is a two roton bound state. Goerbig and Smith have calculated the magneto roton instabilities for higher Landau levels [47].

**D. Bose-Einstein Condensates (BEC).** Gaseous BEC condensates formed from ultra-cold gases of atoms have been predicted to have a roton minimum in the excitations spectrum. The Bijl -Feynman formula, eq. (3.4) can be rewritten for the roton in this case as

$$E(k) = \frac{\hbar^2 k^2}{2m S(k)}$$

where E is the excitation energy and $\hbar k$ the momentum of a Bose liquid. In the BEC case the roton arises through anisotropic dipole-dipole interactions (density-density correlation, eq (3.5)) which are partially attractive; these features are absent in repulsive short-ranged s-wave interactions. O'Dell and others [48] have been able to drive the transition by laser intensity as illustrated by the BEC structure factor $S(k)$, and as further discussed by Minguzzi, March and Tosi [49]. This type of structure is particularly interesting to study as the roton minimum has been suggested to being very close to incipient crystallization as the ordering of atoms leads to a peak in $S(k)$ and ultimately, a solid-fluid transition, as discussed by Nozieres [50]. This provides a connection with the earlier paper, Phase Transition – I [11]. The analogy of this induced BEC with the rest of this section seems clear, especially as a tool to tune phase transition.

**4. Highly Correlated Insulators and Superconductivity**

**A. Insulators.** Peierls recognized the similarities of this phenomenon to superconductivity, i.e. an energy gap opening at the Fermi level [3]. One of the speculations from his work has been the ability to create a traveling charge density wave (CDW), a perfect conductor since a 1D superconductor cannot have a Meissner effect. However, all of the CDW waves created thus far have been "pinned" (localized). Much later, when the organic materials TTF-TCNQ were examined in the 1970's, this effect instead produced an insulating transition.

Su-Schrieffer-Heeger (SSH) [51] used the Peierls' distortion as the underlying basis to describe superconductivity in polyacetylene (PA). They recognized that the Peierls' reorganization energy described above would produce a double well minimum from bond alteration which is associated with spontaneous symmetry breaking resulting in a two fold degenerate ground state. On doping, this system permits nonlinear excitations which are equivalent to moving domain walls which act as topological solitons and give rise to unusual charge and spin states.

We have suggested that an insulating state is present in the doped fullerides [52]; specifically, when $C_{60}$ is doped with two electrons donated by an alkali metal such as potassium, instead of creating an electrical conductor (band theory), the electrons are localized on a fulleride molecule, distort the carbon structure and formed a charge 2e circular CDW which has much stronger binding than a Cooper pair. In essence, this is a highly correlated insulator. The superconducting transition is discussed below.

**B. Superconducting Fullerides [52].** The fulleride insulator in section 4A is the result of the interaction of two CDW's moving in opposite directions in a potential well on the surface of a fulleride. The motion is analogous to Cooper's original argument where one electron motion creates a distortion of the lattice potential which is more attractive than normal for the second electron. In principle the two CDW's are not in perfect phase at higher temperatures, but as the temperature is reduced, the CDW's phase lock, the renormalized vibration frequency goes to zero which defines a transition temperature where a frozen-in distortion occurs. A mean field transition temperature for a 2e charge density wave, $T_{CDW}^{MF}$, can be estimated as

$$k_B T_{CDW}^{MF} = 1.14 \varepsilon_0 e^{-1/\lambda} \qquad (4.1)$$

with $\lambda$, the electron-vibration coupling constant (dimensionless) defined as

$$\lambda = \frac{g^2 n(\varepsilon_F)}{\hbar \omega_{2kF}}$$

The difference in the charge 2e phase locked density wave is that the binding energy is of the order of 0.1 eV instead of the much lower energy of a BCS Cooper pair and the size is much smaller. This circular charge2 density wave (Circ2DW) could be described as a "pseudo-boson". The resulting gap equation above is identical to the BCS equation, but there is no off-diagonal long range order (ODLRO) or phase stiffness as is found in the BCS wave function.

On further doping which averages to three electrons per fulleride molecule, the alkali metal/fulleride system becomes superconducting. It is tempting to suggest that the superconducting transition should have

$$k_B T_{SC}^{MF} = \varepsilon_0(\lambda) e^{-1/g} \qquad (4.2)$$

where $\varepsilon_0$ is the "boson" energy of the charge 2e density wave on a fulleride molecule.

## C. Strongly Coupled BCS Superconductors.

McMillan [53] has derived the following formula for strongly coupled BCS superconductors by empirically fitting data

$$T_c = \frac{\Theta_D}{1.45}\exp-\left\{\frac{1.404(1+\lambda)}{\lambda-\mu^*(1+0.6\lambda)}\right\}$$

Where $\lambda$ is the electron-phonon coupling constant, $\mu$ is the bare Coulomb repulsion, and $\mu^*$ is the renormalized value or Coulomb pseuodopotential. McMillan's formula for the transition temperature takes into consideration that Coulomb repulsion can be reduced by retardation effects and it can be argued that $\mu^*$ and $\lambda$ are both important for $T_c$ since

$$\frac{\lambda}{(1+\lambda)} - \mu^*\frac{1+0.62\lambda}{1+\lambda} \sim \frac{\lambda}{(1+\lambda)} - \mu^*$$

## D. Disordered Superconductors.

Feigel'man and others [54] studied electrons that are near the mobility edge so they can attract each other and form strongly coupled localized Cooper pairs (CP). This then opens a single electron gap at $T_0$ experimentally observed in transport measurements in the insulating state. Below $T_0$ the system can be described by Anderson's $S = \frac{1}{2}$ pseudo-spin theory. The $S_j^z$ components measure the CP occupation numbers and $S_j^\pm$ are pair creation/annihilation operators. Superconductivity results from CP tunneling between states. The tunneling does not compete with the Coulomb repulsion, but with the random energy of the pair in each orbital state resulting in non-zero averages $\langle S^\pm \rangle$ due to off-diagonal coupling of $S_i^- S_j^+ + S_i^+ S_j^-$ in the effective Hamiltonian. The non-zero random averages do compete with the random field in the z direction, the $h_i S_i^z$ term. These localized nearly critical wave functions have large values of the binding energy and strong correlations in both real and energy space resulting in the phase diagram (Fig 1 [54b]) which has three distinct phases: a critical superconducting (SC) phase at $E_F = E_c$, a superconducting state with a pseudogap, and an insulating state. The critical SC is enhanced relative to BCS by an inhomogeneous distribution of the order parameter and density of states while the pseudogap has a two independent energy scales; the first is a SC gap and the second gap characterizes a binding of local energy pairs in an insulating state. The model leads to an explanation of specific features in the scanning tunneling and Andreev spectroscopy

While this fractal pseudo-spin model in similar to the bosonic model in some way, it is distinctly different in three important respects:
1) The energies of the pseudo-gap and collective energy gap are independent.
2) The fractal nature of the eigenstates suggests a large "coordination number" of interacting pseudo-spins away from the SIT.
3) The order parameter distribution in real space is very inhomogeneous.

A virial expansion method can be applied to the pseudo-spin Hamiltonian to determine the full phase diagram from the superconducting transition temperature [eq. (156)] which they write as

$$T_c = \frac{4e^c}{\pi} E_b e^{-1/g^2}$$

with $E_b$ is the upper energy cutoff and c is the Euler constant.

**E. Other Superconductors**.
It seems that most any boson might be capable of mediating the exchange needed between electrons to achieve superconductivity. Proposed examples range from the early efforts by Little who considered dielectric anomalies in dyes molecules (later excitons in the same system) connected via polymer chains [55] through Bardeen's semiconductor excitons [56] to Ruvalds' plasmons [57]. The basic formulas are usually of the type boson energy times an exponential much as the BCS theory where the boson energy is the Debye energy,

$$T_c = \Theta_{boson} \exp\left(-1/g^2\right)$$

Certainly as material fabrication continues to improve and custom syntheses evolve as they have, these examples may become reality.

**6. Discussion**. **a)** Grüneisen Parameter. Renewed interest has been generated in the Grüneisen parameter with the notion that it might be used to detect a control parameter, say, r, induced quantum critical point (QCP) [58, 59]. It is indeed surprising that there is such an unusual sensitivity of thermodynamics on these tuning parameters. The free energy density, $f$, can be expressed as a function of this control parameter and temperature, $f = f(r,T)$, and the sensitivity can be measured by the usual derivatives of the free energy density with respect to r. A QCP tuned by pressure has the following thermal expansion,

$$\alpha = \frac{1}{V}\frac{\partial V}{\partial T}\bigg|_{p,H} = -\frac{1}{V_m}\frac{\partial S}{\partial p}\bigg|_{T,H} = \frac{1}{V_m}\frac{\partial^2 f(r,T)}{\partial p \partial T}$$

$V_m$ being the molar volume. Similar relations hold for other tuning parameters [60]. The Grüneisen parameter has the following relation ($C_p = T\frac{\partial S}{\partial T}\bigg|_p$)

$$\gamma = \frac{\alpha}{C_p} = -\frac{\left(\partial S/\partial p\right)_T}{V_m T \left(\partial S/\partial T\right)_p}$$

In the p-T plane a line of constant entropy can be expressed as

$$dS = \frac{\partial S}{\partial T}\bigg|_p dT + \frac{\partial S}{\partial p}\bigg|_T dp = 0$$

and the Grüneisen parameter is

$$\gamma = \frac{1}{V_m T} \frac{dT}{dp}\bigg|_S$$

a pressure-caloric effect. The isentropes near a quantum phase transition have an accumulation of entropy near a QCP as the system is maximally undecided as to which ground state to choose (Fig 1 in [60]). Approaching a QCP, the characteristic scale of a Fermi liquid $T_0 \approx x^{\upsilon z}$ collapses to zero leading to a temperature divergence of $\alpha$ at a QCP, $\gamma_T \sim 1/T^{1/\upsilon z}$. An example of a magnetic tuning experiment maximizing entropy in $Sr_3Ru_2O_7$ has been discussed [61], along with several other examples. The Grüneisen parameter should change sign at the QCP and the degree of divergence could differentiate between a spin density wave and a local magnetic QCP.

**b). Bound States from the Continuum.** The notion of bound states (including quasi-particles) has an extended history beginning with Breit, and Yukawa and others [62], to Dyson, Feynman, Nambu, and Dancoff [63]. The two-body wave function of Bethe and Salpeter and the resulting integral equation discussed by them represents an important advancement. Gell-Mann and Low [64] then found a method of explicitly expressing the energy relationship using Feynman equations. Certainly the groundwork was laid for Cooper's advance where not only was a two body "Cooper pair" possible, but a general many-body expression was derived where all electrons with $\pm \hbar \omega$ of the Fermi energy participate in the splitting off of the Cooper pair state from the continuum. A very readable account of Cooper's work is in his Nobel acceptance speech [65].

We have listed a number of QP-QP interactions; key questions are: do they all lead to condensation and a resulting macroscopic phase change? Is a quasiparticle necessarily a prelude to a phase change by direct participation or might it only create favorable conditions (electronic liquid crystal "stripes") for other interactions or perhaps something in between such as classical liquid crystal phase changes [66].

**7. Summary, conclusions and proposals for future work. Needs work** We have presented additional cases (relative to [11]) for phase transitions driven by quasi particle interactions. Included are the symmetrical Anderson model and its connection to the Kondo effect and the Curie temperature for a special spin model. Roton-roton interactions can be found in helium-4, FQHE, composite fermions, and BEC systems. Various insulator and superconductor proposals add to the list. Some of the cases seem clear, but others suggest that perhaps additional study could be fruitful. In addition we have discussed the use of the Grüneisen parameter for possibly finding and verifying quantum critical points. Even in this brief study it seems apparent that the microscopic nature of a "classical" phase change contrasted to a quantum quasiparticle phase change (or material microscopic inhomogeneity) seems to demand a better understanding about phase changes, especially in the quantum regime.

Certainly the last statement bears further experimental and theoretical scrutiny, and we anticipate additional work on the theoretical portion where interesting leads and unusual connections suggest a need for a better understanding.

## Summary of Various Quasi-Particle Phase Transitions

| Phase transition | $E_{char}$ | Quasi-particle interaction | Comments/ref |
|---|---|---|---|
| Symmetric Anderson | $\sqrt{\dfrac{2U\Delta}{\pi^2}}$ | $\exp\left(-\dfrac{\pi U}{8\Delta}\right)$ | Connected to Kondo via Schrieffer/Wolff |
| Kondo effect | $(\Gamma\mu)^{1/2}$ | $\exp\left(\dfrac{-2\Gamma}{\pi|\varepsilon_D|}\right)$ | $\varepsilon_D$ energy of impurity orbital; $\Gamma$ impurity level width |
| Curie temperature | $\dfrac{T_c}{J}$ | $1/\exp\left(-2J_0/k_B T\right)$ | |
| Roton-roton interaction | D cutoff energy | $\exp\left(-1/2|\rho_0(K)g_4|\right)$ | [29] |
| Various SC models | Boson energy | Some BCS-like Others not | |
| McMillan | $\Theta_D$ | $\exp-\left\{\dfrac{1.404(1+\lambda)}{\lambda-\mu^*(1+0.6\lambda)}\right\}$ | |
| HTSC with disorder | cutoff energy $E_b$ | $\exp\left(-1/g^2\right)$ | [54] |
| Exciton SC | $\Theta_0$ | $\exp\left(-1/g^2\right)$ | [55] |
| Plasmon SC | $0.7\Theta$ | $\exp\left[-\dfrac{1+\lambda}{\lambda-\mu^*}\right]$ | [56] |
| Fullerides | $\varepsilon_0$ (boson energy) | $\exp\left(-1/g^2\right)$ | [57] |